\documentclass{PoS}
\usepackage[authoryear,square]{natbib}
\bibpunct{(}{)}{;}{a}{}{,}

\title{The SKA view of cool-core clusters: \\
evolution of radio mini-halos and AGN feedback}

\ShortTitle{The SKA view of cool-core clusters}

\author{
\speaker{Myriam Gitti}$^{1,2}$, 
Paolo Tozzi$^3$,
Gianfranco Brunetti$^2$, 
Rossella Cassano$^2$,
Daniele Dallacasa$^{1,2}$,
Alastair Edge$^4$,
Stefano Ettori$^5$,
Luigina Feretti$^2$,
Chiara Ferrari$^6$,
Simona Giacintucci$^{7,8}$,
Gabriele Giovannini$^{1,2}$,
Michael Hogan$^4$,
Tiziana Venturi$^2$
\\ 
$^1$ ~Dipartimento di Fisica e Astronomia - Universit\`a di Bologna, via Ranzani 1, I-40127 Bologna, Italy;
$^2$~INAF - Istituto di Radioastronomia, via Gobetti 101, I-40129 Bologna, Italy;
$^3$~INAF - Osservatorio Astrofisico di Arcetri, Largo E. Fermi 5,
I-50125 Firenze, Italy;
$^4$~Department of Physics, Durham University, Durham, DH1 3LE UK;
$^5$~INAF - Osservatorio Astronomico di Bologna, via Ranzani 1, I-40127 Bologna, Italy;
$^6$~Laboratoire Lagrange, UMR 7293, Universit\'e de Nice Sophia-Antipolis, CNRS, Observatoire de la C\^{o}te d'Azur, 06300 Nice (FR);
$^7$~Department of Astronomy, University of Maryland,  College Park, MD 20742, USA;
$^8$~Joint Space-Science Institute, University of Maryland, College Park, MD, 20742-2421, USA
\\
E-mail: \email{myriam.gitti at unibo.it}
}

\abstract{ In about 70\% of the population of relaxed, cool-core
  galaxy clusters, the brightest cluster galaxy (BCG) is radio loud,
  showing non-thermal radio jets and lobes ejected by the central
  active galactic nucleus (AGN).  In recent years such relativistic
  plasma has been unambiguously shown to interact with the surrounding
  thermal intra-cluster medium (ICM) thanks to spectacular images
  where the lobe radio emission is observed to fill the cavities in
  the X-ray-emitting gas.  This {\it `radio feedback'} phenomenon is
  widespread and is critical to understand the physics of the inner
  regions of galaxy clusters and the properties of the central BCG.
  At the same time, mechanically-powerful AGN are likely to drive
  turbulence in the central ICM which may also play a role for the
  origin of non-thermal emission on cluster-scales.  Diffuse
  non-thermal emission has been observed in a number of cool-core
  clusters in the form of a radio mini-halo surrounding the radio-loud
  BCG on scales comparable to that of the cooling region.  Large
  mini-halo samples are necessary to establish their origin and
  connection with the cluster thermal properties and dynamics,
  especially in light of future X-ray characterization of the cluster
  cores as it is expected by Athena-XIFU.  We show that All-Sky
  reference survey at Band 2 with SKA1 at confusion limit (rms $\sim$2
  $\mu$Jy per beam) has the potential to detect up to $\sim 620$
  mini-halos at redshift $z<0.6$, whereas Deep Tier reference surveys
  at Band 1/2 with SKA1 at sub-arcsec resolution (rms $\sim 0.2 \mu$Jy
  per beam) will allow a complete census of the radio-loud BCGs at any
  redshift down to a 1.4 GHz power of $10^{22} \rm{\, W \, Hz}^{-1}$.
  We further anticipate that SKA2 might detect up to $\sim 1900$ new
  mini-halos at redshift $z<0.6$ and characterize the radio-mode AGN
  feedback in every cluster and group up to redshift $z\sim 1.7$ (the
  highest-$z$ where virialized clusters are currently detected) and
  even beyond, thus providing a complete picture of the feedback
  phenomenon in clusters and its role in shaping the large scale
  structure of the Universe.}

\FullConference{
Advancing Astrophysics with the Square Kilometre Array\\
June 8-13, 2014\\
Giardini Naxos, Sicily, Italy}

\newcommand{\skipthis}[1]{}

\def\ltsim{\raise 2pt \hbox {$<$} \kern-1.1em \lower 4pt \hbox {$\sim$}}
\def\gtsim{\raise 2pt \hbox {$>$} \kern-1.1em \lower 4pt \hbox {$\sim$}}

\begin{document}

\section{Introduction}
\vspace{-0.1in}

The majority of baryons in galaxy clusters are in the form of diffuse
hot plasma, the intra-cluster medium (ICM), heated mostly by
gravitational processes at temperatures in the range 1-10 keV (going
from groups to massive clusters) and visible in the X-ray band thanks
to its thermal Bremsstrahlung emission.  However the catastrophic
cooling of the gas expected at the center of relaxed systems under
these conditions \citep{Fabian-CF_1994} is not observed
\citep[e.g.,][]{Peterson-Fabian_2006}, suggesting that complex
non-gravitational physical processes must be at work in the ICM to
provide smoothly distributed heating on scales of about 100 kpc and
then stop the cooling process.  Understanding the interplay of
gravitational and non-gravitational physics in the ICM and its
interaction with the relativistic plasma ejected by the central active
galactic nucleus (AGN) is key for understanding the growth and
evolution of galaxies and their central black holes, the history of
star formation and the assembly of large-scale structures.  In
particular, the feedback from the central black hole has turned out to
be an essential ingredient that must be taken into account in any
model of galaxy formation and evolution.  The main evidence of the
action of radio-mode AGN feedback is in these {\it `cool-core'} galaxy
clusters, which hold a special place in the entire field of
extragalactic astrophysics.

\begin{figure}[tbp]
\centering 
\includegraphics[width=.48\textwidth,origin=c,angle=0]{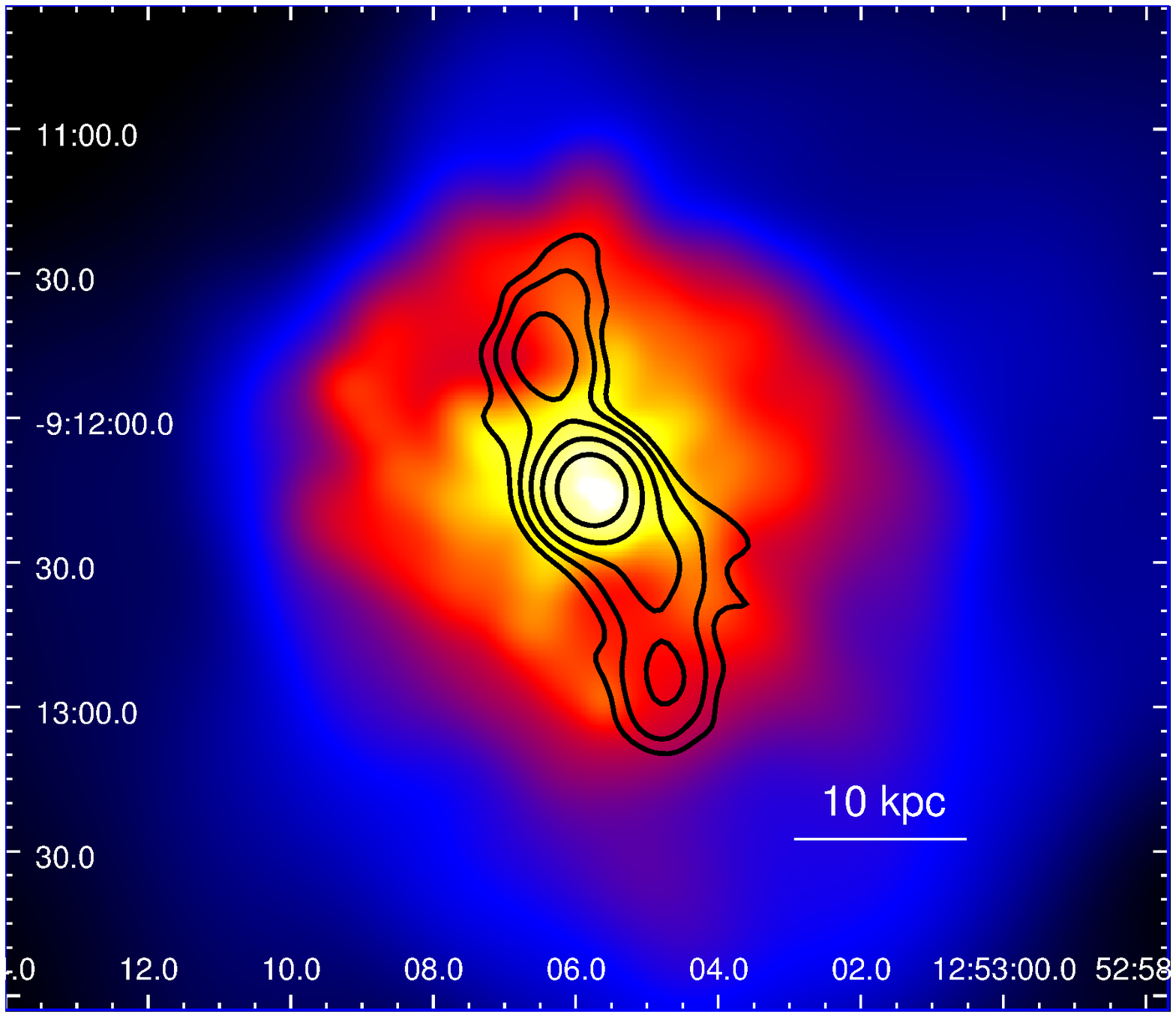}
\hspace{0.4cm}
\includegraphics[width=.48\textwidth,origin=c,angle=0]{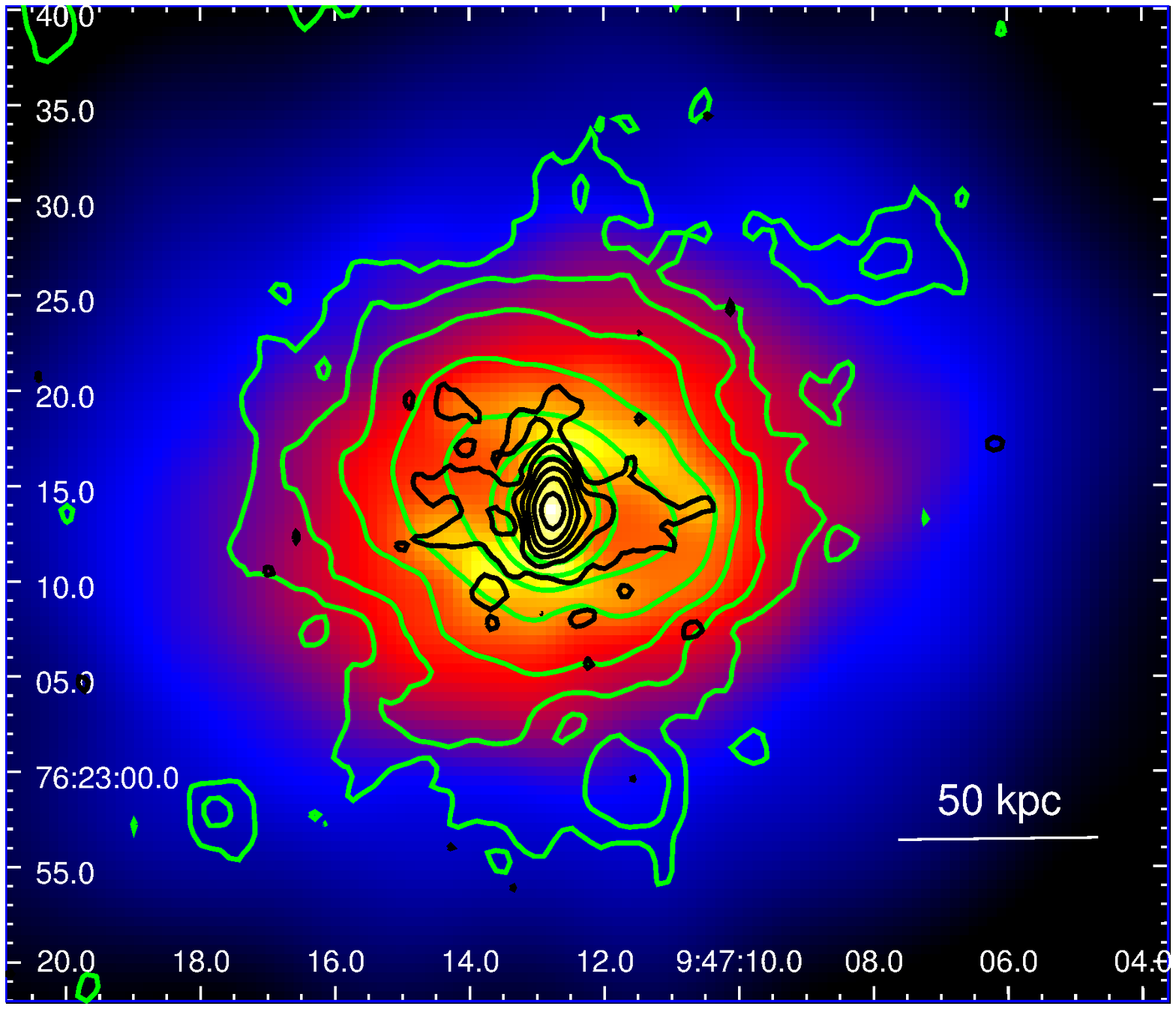}
\vspace{-0.3cm}
\caption{\label{bubbles.fig} {\it Left panel:} X-ray image of the
  galaxy group HCG~62 at $z=0.0137$ \citep{Gitti_2010} with
  superimposed GMRT radio contours at 610 MHz.  {\it Right panel:}
  X-ray image of the galaxy cluster RBS 797 at $z=0.35$ with
  superimposed VLA radio contours at 1.4 GHz \citep[green
  contours,][]{Gitti_2012,Doria_2012} and at 5 GHz \citep[black
  contours,][]{Gitti_2013a}. In both cases X-ray cavities in the ICM
  are observed to be coincident with the BCG radio lobes (black
  contours), and in the case of RBS~797 a large-scale radio mini-halo
  is also detected (green contours).}
\vspace{-0.1in}
\end{figure}

The central dominant (cD) galaxies of cool-core clusters (which are
characterized by short cooling times, high gas densities and low
temperatures in the central regions) have a high incidence of radio
activity, showing the presence of central FR-I radiogalaxies
\citep{Fanaroff-Riley_1974} in 70\% of the cases
\citep{Burns_1990,Dunn-Fabian_2006,Best_2007,Mittal_2009}.
In particular, high-resolution X-ray observations performed with
Chandra and XMM-Newton showed that the central radio sources have a
fundamental and persistent effect on the ICM -- the central hot gas in
many cool-core systems is not smoothly distributed, but shows instead
``holes'' coincident with lobes of extended radio emission. The most
typical configuration is for jets from the central dominant elliptical
of a cluster to extend outwards in a bipolar flow, inflating lobes of
radio-emitting plasma (radio {\it `bubbles'}) which are spatially
separated from the thermal ICM component. The interpretation is that
these lobes push aside the X-ray-emitting gas of the cluster
atmosphere, thus excavating {\it `cavities'} in the ICM which are
detectable as deficits in the X-ray images.  Thanks to spectacular
images where the radio emission is observed to fill the cavities in
the X-ray emitting ICM, the radio galaxies have thus been identified
as a primary source of feedback to solve the so--called `cooling flow
problem' \citep[for recent reviews see][]{Gitti_2012,
  McNamara-Nulsen_2012, Fabian_2012}. In Figure \ref{bubbles.fig} we
show clear examples of radio lobes (black contours) filling the
cavities in the ICM in a galaxy group (left panel) and in a massive
cluster (right panel). 

The detailed mechanism to transfer the mechanical energy of the radio
jets into thermal energy of the ICM is still unclear.  It has been
proposed that mechanically-powerful AGN are responsible for driving
turbulence in the central ICM, which must eventually dissipate into
heat thus contributing to offsetting radiative cooling
\citep[e.g.,][]{Zhuravleva_2014}.  On the other hand, such a
turbulence may also contribute to re-accelerating a seed population of
(sub-)relativistic particles naturally present in the ICM, and
producing diffuse non-thermal emission.  Obvious sources of the seed
relativistic electrons are, for example, the buoyant radio bubbles
that are inflated by the radio activity of the central AGN itself and
disrupted by gas motions in the core.

Diffuse non-thermal emission has been observed in a number of
cool-core clusters, where the radio-loud brightest cluster
galaxy (BCG) is surrounded by a so-called {\it `radio mini-halo'}. 
  Mini-halos are diffuse, faint radio sources observed at the cluster
center on scales (total size) $\sim$100-500 kpc comparable to that of
the cooling region, with steep radio spectra and amorphous (roundish)
shape.  The typical mini-halo power at 1.4 GHz is of the order of
  $\sim 10^{23}$ - $10^{24}$ W Hz$^{-1}$.  These radio sources are
not directly connected with the radio bubbles,
but the mini-halo emission is on larger scales and is truly generated
from the ICM where the thermal plasma and the relativistic electron
population are mixed (see in the right panel of Figure
\ref{bubbles.fig} the example of the cluster RBS~797 which hosts both
a mini-halo, in green contours, and a cavity-bubble system, in black
contours).  Therefore, the synchrotron emission from radio mini-halos
proves that $\sim$GeV electrons are diffusing through $\sim \mu$G
magnetic fields permeating the ICM, as in the case of giant radio
halos and relics \citep[see also][]{Cassano-SKA, Ferrari-SKA,
  Govoni-SKA}.  Mini-halos are qualitatively similar to the most
extreme case of giant radio halos and relics that however are
  tipically found in disturbed/non cool-core clusters and are most
likely powered by cluster mergers.
In fact, radio mini-halos indicate that, in addition to mergers, diffuse
non-thermal emission can be powered also by less extreme processes
that dissipate energy associated to large-scale (tens-hundreds kpc)
motions at micro-physical scales and are connected at some level with
the mechanisms responsible for the heating of the cooling gas.
Although mini-halos are not directly sustained by the central engine,
it is nonetheless likely that the AGN plays a role for the
origin of these sources.  For example, it has been proposed that the
relativistic electrons that generate mini-halos are released in the
ICM by the central AGN and then can be transported and re-accelerated
by turbulent motions in the cool cores \citep{Gitti_2002,
  Mazzotta-Giacintucci_2008, ZuHone_2013}.

Only about twenty mini-halos (or candidates) are known so far, all at
redshift $z<0.6$ \citep{Giacintucci_2014a, vanWeeren_2014}. Due to the
limitations of current radio interferometers it is difficult to assess
the occurrence of radio mini-halos in clusters. In particular, the
mini-halo detection is complicated by the need of separating their
diffuse, low surface brightness emission (at a level of few $\mu$Jy
arcsec$^{-2}$ at 1.4 GHz) from the bright emission of the central
radio BCG. This requires a very good sensitivity to diffuse emission,
high dynamic range and good spatial resolution, which will be
achievable with SKA only.  A large mini-halo sample observed by SKA
will allow us to reach a better understanding of this class of
sources, in particular to establish their origin and connection with
the cluster thermal properties and dynamics which will be
characterized at unprecedented levels by the next-generation of X-ray
instruments like XIFU, the X-ray Integral Field Unit covering the 0.2
to 10 keV energy range with unrivalled energy resolution, that will
flight onboard of the ESA-L2 mission
Athena\footnote{http://www.the-athena-x-ray-observatory.eu/} (expected
launch in 2028).
Furthermore, at present the study of the radio-mode feedback in
clusters is limited to $z < 0.7$ with very few cases at high redshift
\citep{Hlavacek_2012}, while virialized clusters are currently
detected up to redshifts of 1.6 or larger (at present the highest
redshift, massive cluster whose virialization is confirmed by deep
X-ray study is XDCP J0044.0$-$2033 at $z=1.58$, Tozzi et al.,
submitted). Therefore, there is a wide redshift range where radio-mode
feedback is still unexplored.  In addition, recent studies show that
cool cores are already present and well developed at least at $z\sim
1$, suggesting that cool-core formation takes place on a short
time-scale after the formation of the cluster \citep{Santos_2010,
  Santos_2012}.
Since the association of cool cores with radio mini-halos and radio
AGN is ubiquitous \citep{Feretti_2012, Sun_2009b}, with SKA we expect
to be able to trace the non-thermal emission and feedback activity in
clusters, and therefore to study the interplay between thermal and
non-thermal plasma in the central regions of galaxy clusters, up to
the highest redshift where clusters are found.

The non-thermal activity and radio-mode feedback in clusters of
galaxies is a scientific case key to the physics of the ICM and the
evolution of the cosmic large scale structures.  Radio astronomy in
the SKA era can be the main scientific driver in these fields.

In the rest of this chapter we assume a $\Lambda$CDM cosmology with
$H_0 = 70 \mbox{ km s}^{-1} \mbox{ Mpc}^{-1}$ and
$\Omega_M=1-\Omega_{\Lambda}=0.3$.

\section{Origin of radio mini-halos}
\label{origin.sec}
\vspace{-0.1in}

At a first glance, radio mini-halos may resemble a small-scale version
of giant radio halos, extending on scales of the order of few hundred
kpc size rather than Mpc size. However, the two classes of radio
sources show prominent differences. First of all, in clear contrast to
giant radio halos\footnote{note that there are also two atypical cases
  of giant halos, that apparently are in clusters without strong
  merging activity \citep{Farnsworth_2013,Bonafede_2014}}, mini-halos
are always found in dynamically relaxed systems suggesting that
cluster mergers do not play a major role for their origin. Also the
synchrotron volume emissivity of mini-halos is typically larger than
that of giant halos \citep{Cassano-Gitti_2008, Murgia_2009}.  On the
other hand, it is yet not well established whether the underlying
physical mechanisms that accelerate the cosmic ray electrons (CRe) in
mini-halos differ substantially from the analogous mechanisms in giant
radio halos. Clusters hosting mini-halos always have a central
radio-loud AGN, which often exhibits outflows in the form of radio
lobes and bubbles injecting CRe into the central regions of galaxy
clusters.  In principle these AGN could represent the primary source
of the CRs in mini-halos, however they are not sufficient by
themselves, at least without some dynamical contribution, to explain
the diffuse radio emission. In particular, a slow diffusion problem
exists for mini-halos (as for giant halos); that is, the energy loss
time scale of the radio emitting CRe ($\approx 10^8$ yr) is much
shorter than the time needed by these particles to diffuse efficiently
across the emitting volume \citep[e.g.,][]{Gitti_2004}.
Similarly to giant halos, two physical mechanisms have been identified
as possibly responsible for the radio emission in mini-halos: i)
re-acceleration of CRe ({\it leptonic} models or re-acceleration
models) where electrons are reaccelerated by micro-turbulence in the
radio-emitting region, and ii) generation of secondary CRe ({\it
  hadronic} or secondary models) where the radio-emitting region is
continuously supplied with secondary electrons generated by inelastic
collisions between cosmic-ray protons (CRp) and thermal protons
\citep[see][for a review]{Brunetti-Jones_2014}.

A key question in the re-acceleration models is the origin of the
turbulence responsible for re-accelerating the electrons.
\cite{Gitti_2002} originally considered magnetohydrodynamic (MHD)
turbulence amplified by compression in the cool cores, where the
necessary energetics to power radio mini-halos is supplied by the
cooling flow process itself.  This model predicts a direct connection
between the synchrotron emitted power in mini-halos and the energetics
of (or energy available into) the cooling flow.  Remarkably a trend
between the radio power and the cooling flow power is observed in
mini-halo clusters \citep{Gitti_2004, Gitti_2007b, Gitti_2012}, where
the strongest radio mini--halos are associated with the most powerful
cooling flows, thus hinting at a direct connection between the thermal
properties of cooling flows and the non-thermal emission of radio
mini-halos.
Although these observations support a connection between synchrotron
and (gas-)cooling power, the physical mechanisms responsible for the
acceleration of the emitting electrons are still poorly
constrained. In particular, in the re-acceleration model the origin of
the turbulence necessary to trigger the electron re-acceleration is
still debated.  As already mentioned in the Introduction, one
possibility is that the turbulence is powered by AGN activity. In this
case the mechanisms responsible for particles re-acceleration in
mini-halos and for the gas heating should be intimately connected.
This scenario can be tested by investigating whether all mini-halo
clusters show evidence of AGN feedback (see Sect. \ref{feedback.sec}).
On the other hand, the signatures of minor dynamical activity have
recently been detected in some mini-halo clusters, thus suggesting
that additional or alternative turbulent energy may be provided by
minor mergers \citep{Gitti_2007b, Cassano-Gitti_2008} and related gas
sloshing mechanism in cool-core clusters
\citep{Mazzotta-Giacintucci_2008, Giacintucci_2014b}, which is also
supported by recent simulations \citep{ZuHone_2013}.
The clearest observational signatures of these large-scale gas motions
are spiral-shaped cold fronts seen in the majority of cool-core
clusters \citep{Markevitch-Vikhlinin_2007}. These cold fronts are
believed to be produced by the cold gas of the core sloshing in the
clusters deep potential well, in response to crossing dark matter
subhalo motions, for example. Those sloshing motions can advect ICM
across the cluster core where turbulence can also be produced
\citep{Fujita_2004, Ascasibar_2006, ZuHone_2010}.

Understanding the nature of mini-halos would provide invaluable
information on the micro-physics of the ICM, including the process of
amplification and dissipation of the magnetic field, and on the
diffusion/transport of CRs in these environments.  Both understanding
the nature of turbulence in cool-core clusters and discriminating
between a leptonic and hadronic origin of radio mini-halos is however
very challenging due to severe limitations of current observational
constraints.
A leap forward is expected from future spectral and polarimetric
information in the radio band, that will become available with SKA,
combined with constraints in other bands (e.g., hard X-rays and
gamma-rays).  For example, the combination of SKA1-LOW and
SKA1-MID/SUR observations will allow us to establish whether the
spectrum of mini-halos is curved or if it is a pure
power-law. Compelling evidence for very steep spectra and/or for the
existence of spectral breaks at high frequencies would favour a
re-acceleration scenario rather than a hadronic origin of the mini
halos \citep{Brunetti-Jones_2014}. At the same time, polarimetric
information will allow us to probe the turbulent status of the
magnetic field in the radio-emitting region, whereas the degree of
circular polarization will allow us to poinpoint the CRe to CRp ratio,
thus constraining the hadronic or re-acceleration origin of radio
mini-halos.

In this chapter we focus on the crucial information that will be
obtained by statistical studies with SKA.  Although a correspondence
between radio mini-halos and cool-core clusters is well established,
current radio studies do not provide an exhaustive view of the
occurrence of radio mini-halos in galaxy clusters more generally. For
example, it is still not clear whether these radio sources are common
or rare in cool cores.
Statistical studies of large cluster samples are necessary to reach a
better understanding of these sources, investigating in particular the
following key points: \vspace{-0.4cm}
\begin{itemize}
\item Do all cool-core clusters host a radio mini-halo?  How does the
  mini-halo/cool-core fraction evolve with redshift?
  How do the radio properties correlate with the X-ray properties? \\
  {\it (radio power-limited sample with wider redshift
    distribution, synergy with eROSITA and Athena X-ray satellites)}
  \vspace{-0.3cm}
\item 
Does the central AGN play a significant role in powering radio mini-halos?
What is the fraction of mini-halo clusters that show evidence of AGN 
feedback? \\
{\it (spectral studies, radio bubbles filling the X-ray cavities)} 
\vspace{-0.3cm}
\item Are mini-halos intrinsically different from giant halos, or do
  they represent a different evolutionary stage of the same
  non-thermal phenomenon?  In particular, if non cool-core clusters
  evolve into cool-core clusters as believed, do giant radio halos
  evolve into
  mini-halos?\\
  {\it (polarimetric studies, evolutive models and synergy with Athena
    X-ray satellite)}
\end{itemize}
\vspace{-0.1cm} We will present evidence that surveys with SKA1 and
eventually with SKA2 will have the capabilities to address the above
questions.

\section{Statistics of radio mini-halos}
\label{statistics.sec}
\vspace{-0.1in}

We start from the list of mini-halos reported in
\cite{Giacintucci_2014a} who recently selected a large sample of
X-ray-luminous clusters with available high-quality radio data, and
discovered four new mini-halos. By excluding the objects that they
classify as ``candidate'' or ``uncertain'', and by including the new
detection in the Phoenix cluster \citep{vanWeeren_2014}, the mini-halo
sample used in this work comprises the following 16 objects (sorted by
decreasing radio power at 1.4 GHz):
RX~J1347.5$-$1145 (z=0.451), 
Phoenix (z=0.596), 
RX~J1720.1$+$2638 (z=0.159),
A~2390 (z=0.23), 
RX~J1532.9$+$3021 (z=0.362), 
RXC~J1504.1$-$0248 (z=0.215),
RBS~797 (z=0.35), 
Perseus (z=0.018),
MS~1455.0$+$2232 (z=0.258), 
ZwCl~3146 (z=0.290),
A~1835 (z=0.252), 
A~2204 (z=0.152), 
A~478 (z=0.088), 
A~2029 (z=0.077), 
Ophiuchus (z=0.028), 
2A~0335$+$096 (z=0.035).
We note that all these mini-halo clusters have a central
entropy $K_0 = kT_0 n_0^{-2/3}$ \ltsim 25 keV  cm$^2$ \citep[where $K0$ values are
taken from the  Chandra ACCEPT\footnote{ Archive of
  Chandra Cluster Entropy Profile Tables} sample of][]{Cavagnolo_2009}, that according
to the classification of \cite{Hudson_2010} defines the
population of strong cool cores (SCC).
Recently, \cite{Panagoulia_2014a} take issues with the ACCEPT sample,
arguing that the flattening of the entropy profile towards smaller
radii, that is, the measured excess `floor' $K_0$ of the core entropy
above the best-fitting power-law profile for the entropy at larger
radii, could be a resolution effect.  Since the measured central
entropy is correlated with the central bin size, this problem is
enhanced for clusters with poor data quality, or poor spatial
resolution, where the inner regions are undersampled. However, we note
that this effect is not an issue in this context, as it simply implies
that the true central entropy value could be even lower, thus
strengthening the SCC classification of our mini-halo sample (see also
Giacintucci et al. in preparation for a more detailed investigation).

\begin{figure*}[t]
\parbox{0.5\textwidth}{
\centerline{\includegraphics[scale=0.38, angle=0]{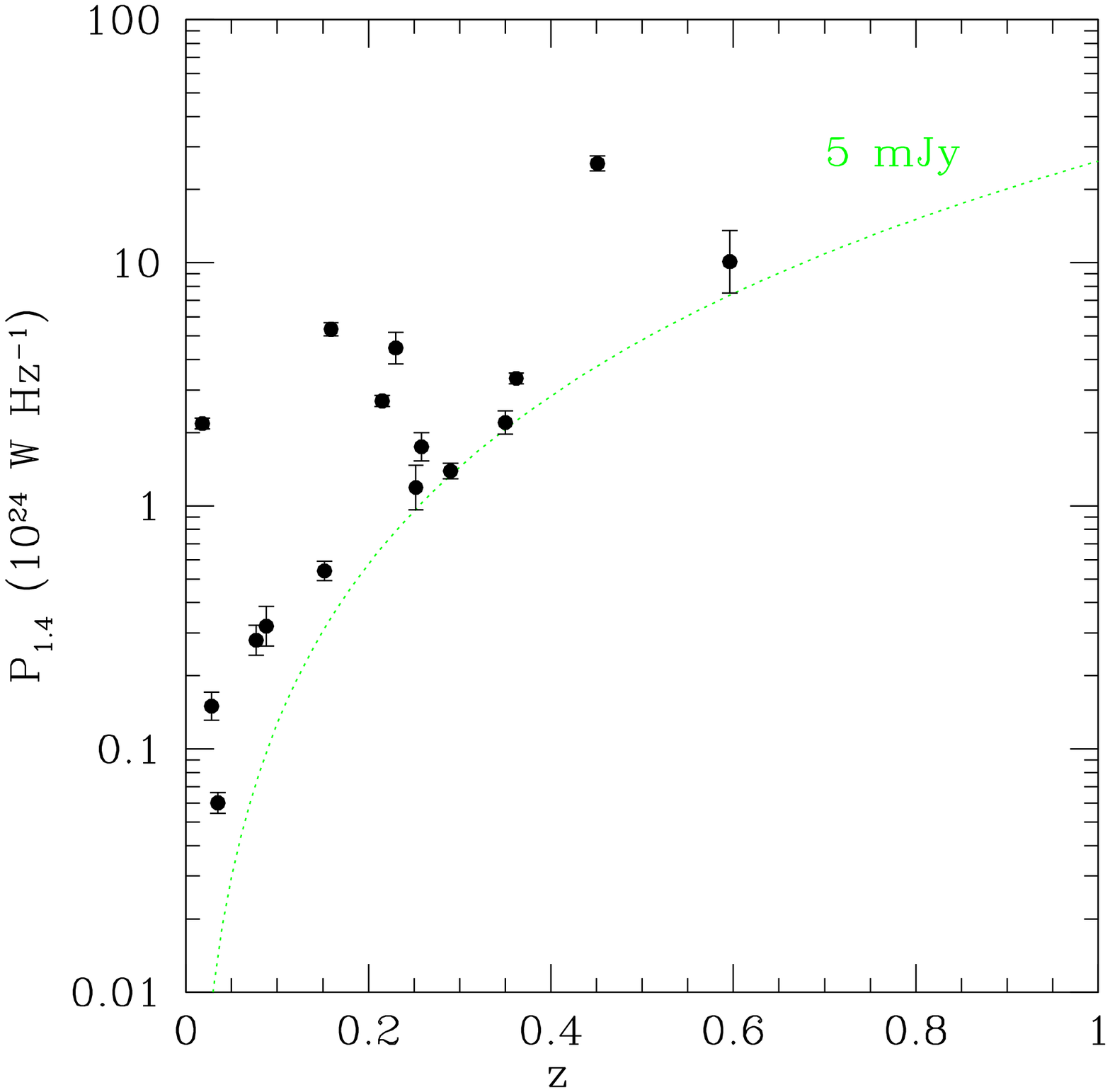}}
}
\hspace{0.4cm}
\parbox{0.5\textwidth}{
\centerline{\includegraphics[scale=0.38]{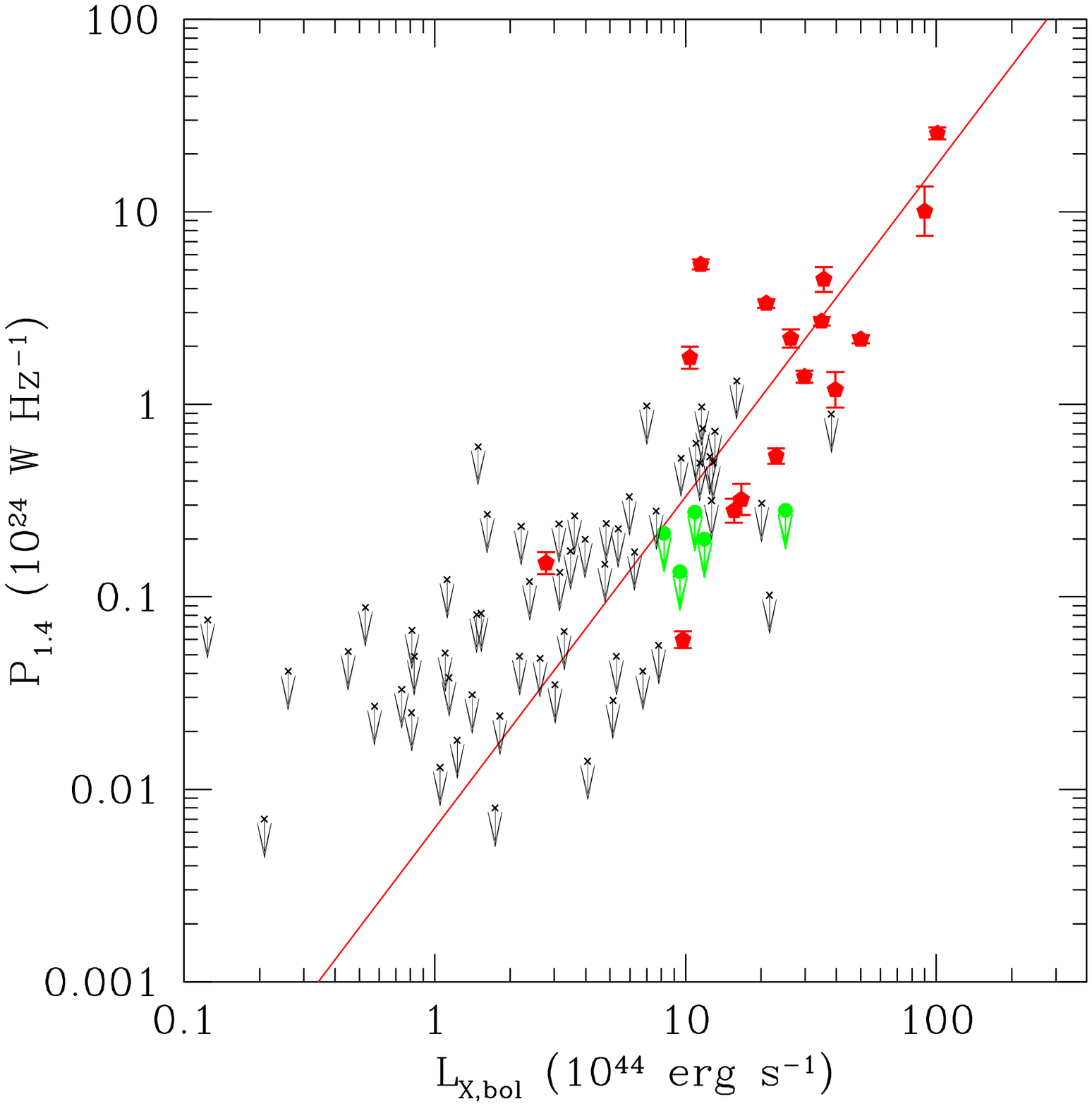}}
}
\vspace{-0.1cm}
\caption{\label{statistics.fig}{\it Left panel:} Radio power at 1.4
  GHz versus redshift for the mini-halo sample. The green dotted line
  represents the radio power corresponding to a flux density of 5
  mJy. {\it Right panel:} Radio power at 1.4 GHz versus the
  cool-core-excised bolometric X-ray luminosity for the mini-halo
  sample (red filled circles). The red solid line is the best fit
  relation to the red points from bisector BCES regression to the parameters in log
  space (see Eq. 3.1).  The green points are the observed upper limits
  from \cite{Kale_2013}, whereas the black crosses are the current upper
  limits estimated for the ACCEPT clusters which are candidates to
  host radio mini-halos.  }
\vspace{-0.1in}
\end{figure*}

In Figure \ref{statistics.fig} (left panel) we plot the 1.4 GHz radio
power of the mini-halos \citep[taken from][]{Giacintucci_2014a,
  vanWeeren_2014} versus redshift. For comparison we overlay the radio
power corresponding to a flux of 5 mJy (green dotted line), which is
roughly the lowest mini-halo flux detected up to now.  It is evident
that there is a strong observational bias that limits our present
ability of detecting mini-halos; that is, we are currently missing
many faint mini-halos simply because of the limited sensitivity of
present radio telescopes.  But how many mini-halos await discovery?
The technical improvement reachable with SKA will be fundamental to
answer to this question.
To estimate the quantum leap produced by SKA in the ability of
detecting radio mini-halos, we must link the non-thermal properties of
these sources with the thermal properties of their host galaxy
clusters.  Being dominated by selection effects, the current radio
information on mini-halos does not allow us to make predictions on
future discoveries.  By contrast, cluster statistics in terms of X-ray
properties is already available from Chandra and XMM studies, and can
be exploited to forecast future detections of radio mini-halos,
provided an intrinsic relation between the thermal and non-thermal
cluster properties exists.

To link the properties of the mini-halos with the global X-ray
properties of the host clusters, in Figure \ref{statistics.fig} (right
panel) we plot with red filled circles the mini-halo radio power at
1.4 GHz versus the cool-core-excised bolometric X-ray luminosity of
the cluster, taken from the ACCEPT sample.  We used the bivariate
correlated error and intrinsic scatter (BCES) algorithm
\citep{Akritas_1996} to perform regression fits to the data in log
space, determining the best-fitting powerlaw relationship (bisector
method) between the radio power, $P_{1.4}$, and the bolometric X-ray
luminosity, $L_{\rm X,bol}$ (overlaid as a red solid line in Figure
\ref{statistics.fig}, right panel):
\begin{equation}
\log P_{1.4} = 1.72(\pm 0.28) \, \log L_{\rm X,bol} - 2.20(\pm 0.46)
\label{bestfit}
\end{equation}
\noindent where $P_{1.4}$ is in units of $10^{24}$ W Hz$^{-1}$ and
$L_{\rm X,bol}$ is in units of $10^{44}$ erg s$^{-1}$.
We note that our analysis improve significantly the previous ones
\citep[e.g.,][]{Cassano-Gitti_2008, Kale_2013} thanks to the better
statistics, and that a similar scaling (with slope $1.76\pm0.16$) is
obtained for giant radio halos \citep{Brunetti_2009}.  It is also
entirely plausible that the flux density of mini-halos scales, perhaps
even better, with other X-ray properties as well, such the cool-core
strength\footnote{a detailed investigation of this correlation, which
  requires accurate model fits to the X-ray spectra extracted inside
  the cooling region, is currently underway and will be presented in a
  forthcoming paper}.  However, we stress that for statistical studies
of large cluster samples it is essential to link the non-thermal
properties of the mini-halos to the thermal properties of the ICM
which are easily observable with X-ray surveys, such as the X-ray
luminosity.

Our starting point is the {\it optimistic} zero-th order assumption
that every SCC cluster hosts a radio mini-halo, and that such a
mini-halo follows the radio--X-ray power correlation.  To check
whether our hypothesis is consistent with the current independent
constraints, in Figure \ref{statistics.fig} (right panel) we report in
green the observed upper limits to the mini-halo emission in 5 SCC
clusters from \cite{Kale_2013}.  Although these limits appear
systematically lower than the correlation, they are still consistent
with the scatter of the observed trend.  In Figure
\ref{statistics.fig} (right panel) we also report the radio upper
limit distribution of all SCC clusters in the ACCEPT sample which are
candidate to host radio mini-halos; that is, according to our
assumption, clusters having a central entropy $K0$ \ltsim 25 keV
cm$^2$.
Although most of these clusters have radio observations, there is no
evidence for the presence of mini-halos in these clusters
\citep[see][]{Giacintucci_2014a}.  Thus a mandatory point is to check
whether the absence of mini-halos in these SCC clusters is still
consistent with our assumptions. We did that by estimating the radio
upper limits in Figure \ref{statistics.fig} (right panel) based on the
capabilities of current radio telescopes. In particular we adapted to
mini-halos the criterion of \cite{Cassano_2012} based on a threshold
in flux for giant halos.  From this threshold we derive the minimum
flux of radio mini-halos that can be detected by assuming a spatial
distribution of their brightness, and use this value to derive the
upper limit to the luminosity of mini-halos in each SCC cluster.
\cite{Murgia_2009} modeled the radio brightness profile $I(r)$ of
radio mini-halos with an exponential of the form $I(r) = I_0
\exp(-r/r_e)$, where $r_e$ is the effective radius.  This implies that
the outermost, low brightness, regions of mini-halos are very
difficult to detect.  However, what is important is the capability to
detect at least the brightest mini-halo regions.  By considering the
brightness profile of \cite{Murgia_2009}, we estimate that radio
mini-halos emit about half of their total radio flux within their half
radius $r_{50} \sim 1.68 \, r_e$.  Following \cite{Cassano-SKA} we can
derive the minimum detectable flux, $f_{\rm min}(z)$, by assuming that
the mini-halo is detected when the integrated flux within $r_{50}$
gives a signal to noise ratio $\xi_2$.
In particular, for each low-entropy cluster selected in the ACCEPT
sample, we estimate $f_{\rm min}(z)$ from Eq. 4.2 of
\cite{Cassano-SKA} by considering the angular size of the radio
mini-halo in arcseconds, $\theta_{\rm MH}(z)$, which corresponds to $2
\, r_{50}$ at the given cluster redshift. We adopt typical values of
the current sample of observed mini-halos
\citep[e.g.,][]{Giacintucci_2014a} of half radius $r_{50} \sim 100$
kpc, rms noise per beam F$_{\rm rms}$ = 25 $\mu$Jy, $\xi_2 = 10$, beam
angular size $\theta_b = 10$ arcsec.
The observed (green points) and estimated (black crosses) upper limits
in Figure \ref{statistics.fig} (right panel) do not violate the
$P_{1.4}$-$L_{\rm X,bol}$ correlation, therefore we shall proceed with
the assumption that all SCC clusters host a mini-halo that follows the
radio--X-ray power correlation.

\begin{figure*}[t]
\parbox{0.5\textwidth}{
\centerline{\includegraphics[scale=0.38]{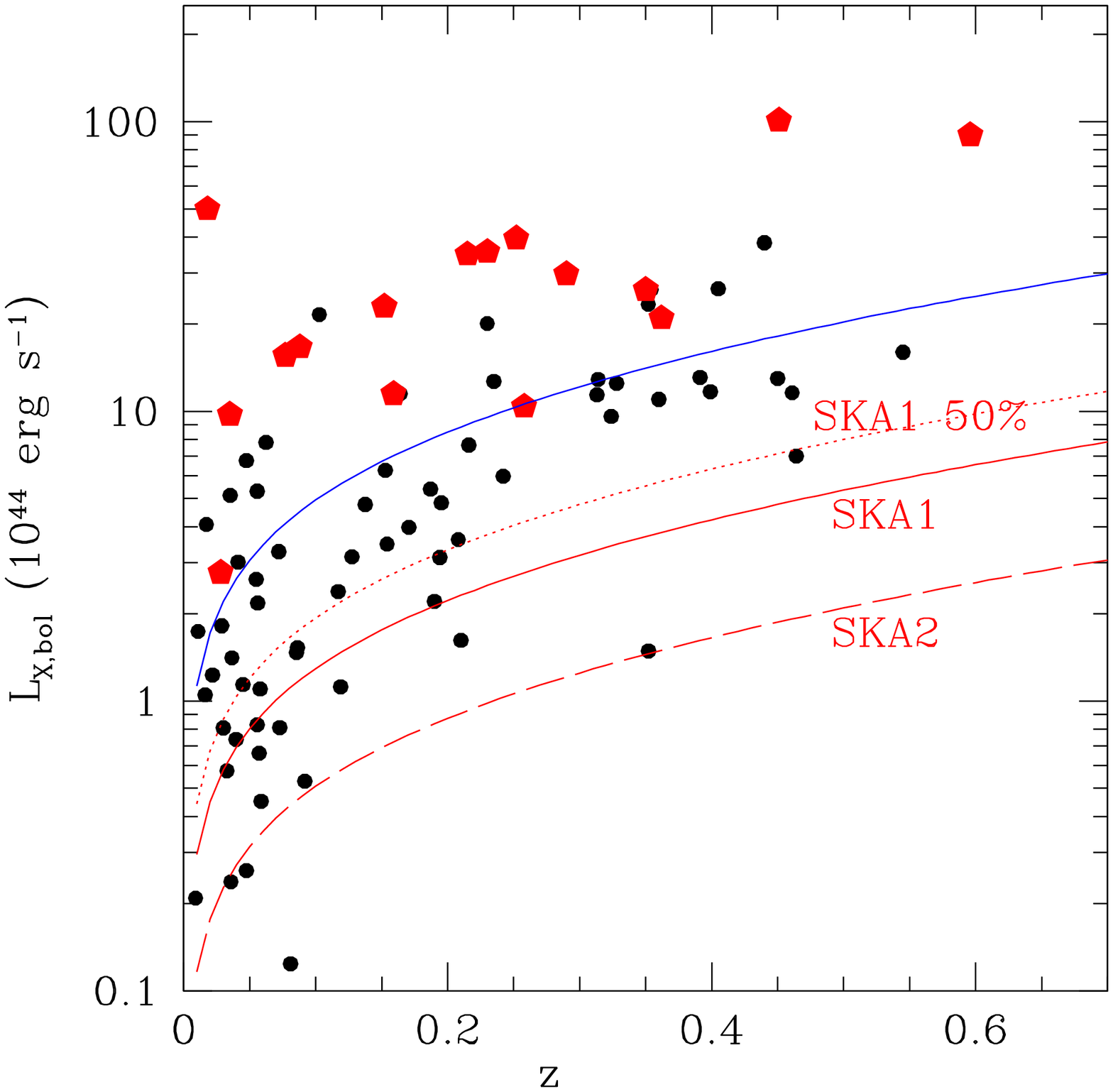}}
}
\hspace{0.4cm}
\parbox{0.5\textwidth}{
\centerline{\includegraphics[scale=0.38]{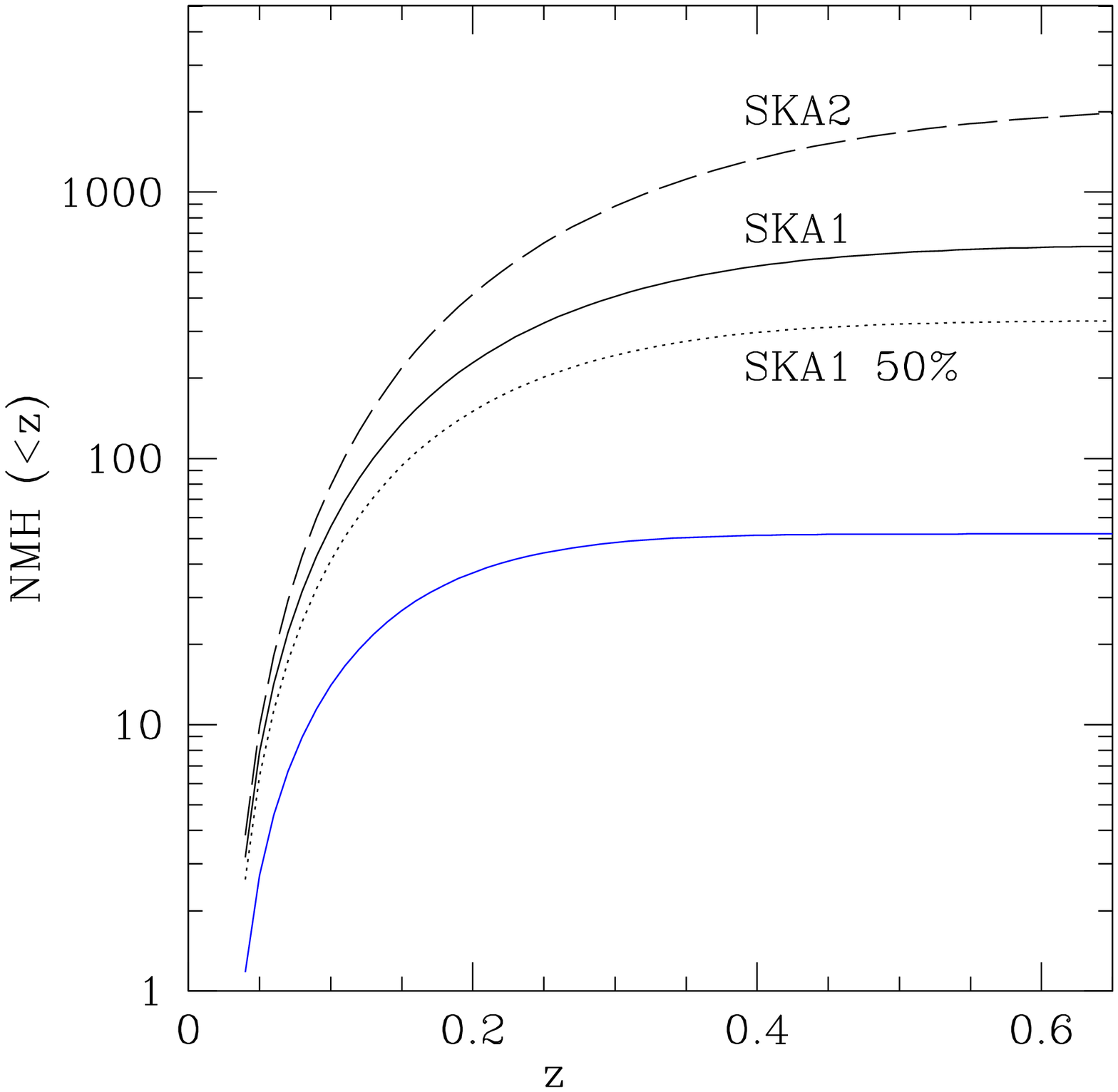}}
}
\vspace{-0.1cm}
\caption{\label{predictions.fig}{\it Left panel:} Cool-core-excised
  bolometric X-ray luminosity, $L_{\rm X,bol}$, versus redshift for the ACCEPT
  clusters with central entropy $K0$ \ltsim 25 keV cm$^2$
  \citep{Cavagnolo_2009}. The clusters hosting a radio mini-halo are
  highlighted in red.  The blue solid line is indicative of the
  current mini-halo detection limit on the population of SCC clusters
  (F$_{\rm rms}$ = 25 $\mu$Jy, $\theta_b = 10$ arcsec), whereas the red solid
  line represents the detection limit reachable by SKA1-SUR surveys at
  confusion limit (obtained from the minimum radio power calculated
  for observations at 1.4 GHz with F$_{\rm rms}$ = 2 $\mu$Jy, $\theta_b = 8$
  arcsec). The red dotted lines and red dashed lines represent the
  detection limit reachable by early science operations of
  SKA1-SUR (F$_{\rm rms}$ = 4 $\mu$Jy, $\theta_b = 8$ arcsec), and by SKA2 (rms
  = 0.2 $\mu$Jy, $\theta_b = 4$ arcsec), respectively.
  {\it Right panel:} Integrated number of radio mini-halos detectable
  at 1.4 GHz out to $z$, NMH ($< z$), as a function of redshift.  The
  predictions for All Sky surveys with SKA1, early SKA1 and SKA2 are
  shown with black solid lines, black dotted lines and black dashed
  lines, respectively. The prediction for a hypothetical survey
  conducted with current telescopes is also shown for comparison (blue
  solid line). The survey performances adopted to estimate the
  integrated number of mini-halos are the same as in left panel. 
    The uncertainty envelopes driven by the $P_{1.4}$-$L_{\rm
   X}$ correlation alone are of the order of $\sim$30\% at
      z=0.1, $\sim$40\% at z=0.3 and $\sim$50\% at z=0.6, but are not
      shown here to avoid confusion among different lines.  
\vspace{-0.1in} }
\end{figure*}

To derive a zeroth-order estimate of the detection limit reachable by
SKA, in Figure \ref{predictions.fig} (left panel) we plot the clusters
in the ACCEPT sample which are candidates to host radio mini-halos
(i.e., having $K0$ \ltsim 25 keV cm$^2$) as a function of redshift. In
red we highlight the objects that are known to possess a mini-halo,
and in blue overlay the threshold in X-ray bolometric luminosity
obtained by converting the upper limits on $P_{1.4}$ (black points in
Figure \ref{statistics.fig}, right panel) to upper limits on $L_{\rm
  X,bol}$ by means of Eq. \ref{bestfit}.  The blue line thus is
representative of the current mini-halo detection limit on the
population of SCC clusters.  As clear, at present we are able to
investigate only the ``tip of the iceberg'' of the population of SCC
clusters.  We then estimate the minimum flux as described above by
assuming values in the reach of SKA1-SUR surveys at confusion limit
\citep[see][for more discussion on the role of
confusion]{Prandoni-SKA,Cassano-SKA}.  In particular, the red solid
line in Figure \ref{predictions.fig} (left panel) represents the
minimum X-ray bolometric luminosity obtained (via the
$P_{1.4}$-$L_{\rm X,bol}$ correlation given by Eq. \ref{bestfit}) from
the minimum 1.4 GHz radio power calculated for observations at 1.4 GHz
with F$_{\rm rms}$ = 2 $\mu$Jy, $\xi_2 = 10$, $\theta_b = 8$ arcsec.
We also estimate that the scientific outcomes delivered in this field
during early science operations of SKA1 (with performances at
$\sim$50\%) will be significant, allowing the radio follow-up to more
than 70\% of the ACCEPT sample (see red dotted line in Figure
\ref{predictions.fig}, left panel).  Our anticipations for SKA2 are
shown in Figure \ref{predictions.fig}, left panel (red dashed line).

\subsection{Number of radio mini-halos expected in surveys with SKA1-SUR}

Given the large number of stations and the unprecedented {\it (u,v)}
coverage of SKA, the detection will depend only on the mini-halo flux
density.  The number of mini-halos that can be detected from a radio
survey with a given sensitivity up to a certain redshift $z$ is
computed by integrating the radio luminosity function (RLF) of
mini-halos $ {{d N_{\rm MH} }\over{dP_{1.4} dV}} $ over radio
luminosity and redshift \citep[cf. Eq. 5.1 of][]{Cassano-SKA}, where
the minimum radio luminosity detectable at a given redshift can be
estimated from $f_{\rm min}(z)$ as described above \citep[Eq. 4.2
of][]{Cassano-SKA}.  Under our assumptions the RLF of mini-halos (per
sky area surveyed in steradians) is
\begin{equation}
 {{d N_{\rm MH} }\over{dP dV}} = {\rm f}_{\rm SCC} \, {{d N_{\rm cl}
  }\over{dL_{\rm X} dV}} \, {{dL_{\rm X}} \over{dP_{1.4}}} 
\label{rlf.eq}
\end{equation} 
\noindent
where ${{d N_{\rm cl} }\over{dL_{\rm X} dV}}$ is the X-ray luminosity
function (XLF) of galaxy clusters
\citep[e.g.,][]{Mullis_2004,Bohringer_2014}, ${\rm f}_{\rm SCC}$ is
the fraction of clusters with SCC
\citep[e.g.,][]{Hudson_2010,Bharadwaj_2014} and ${dL_{\rm
    X}}\over{dP_{1.4}}$ can be obtained from the observed radio--X-ray
power correlation for mini-halos.

In particular, here we adopt the evolving XLF derived from the
high-redshift, X-ray--selected 160 Square Degree {\it ROSAT} Cluster
Survey (160SD) by \cite{Mullis_2004} in a $\Lambda$-dominated
universe, assuming the local XLF determined from the REFLEX survey
\citep{Bohringer_2002}.  For consistency with the 160SD, where $L_{\rm
  X}$ is calculated in the 0.5-2.0 keV band, we derive the
$P_{1.4}$-$L_{\rm X}$ correlation for mini-halos by estimating $L_{\rm
  X}$ in the same energy band. For each cluster in our mini-halo
sample, we take the X-ray luminosity from the meta-catalogue of
\cite{Piffaretti_2011}, which is given in the 0.1-2.4 keV energy band
inside the radius $R_{500}$, and convert it to the 0.5-2.0 keV energy
band assuming an Xspec {\ttfamily mekal} plasma model at the observed
cluster temperature \citep[taken from][]{Cavagnolo_2009}, redshift,
and assuming a metallicity of 0.3. We derive a correlation in the form
$ \log P_{1.4} = 2.03(\pm 0.20) \, \log L_{\rm X} - 1.65(\pm 0.21) $,
where $P_{1.4}$ is in units of $10^{24}$ W Hz$^{-1}$ and $L_{\rm X}$
is in units of $10^{44}$ erg s$^{-1}$.  Hence in Eq. \ref{rlf.eq} we
can substitute
\begin{equation}
{{dL_{\rm X}}\over{dP_{1.4}}} = {{L_{\rm X}}\over{P_{1.4}}} {{d \log  L_{\rm
      X}}\over{d \log P_{1.4}}} = {{1}\over{2.03}} {{L_{\rm X}}\over{P_{1.4}}}
\end{equation}
\noindent 
By further assuming ${\rm f}_{\rm SCC} \sim 0.40$ \citep{Hudson_2010},
we estimate the number counts of radio mini-halos as a function of
redshift from the All Sky surveys ($3 \pi$) with SKA1, early SKA1 and
SKA2 (see Fig. \ref{predictions.fig}, right panel); we predict that
they will be able to detect up to $\sim 620$, $\sim 330$ and $\sim
1900$ mini-halos, respectively, out to redshift $z \sim 0.6$.  If the
cluster XLF is taken as known, a measure of the error on such
estimates can be obtained by considering the combined maximum
variation of the $P_{1.4}$-$L_{\rm X}$ correlation parameters. This
leads to uncertainties of the order of $\sim$30\%, $\sim$40\% and
$\sim$50\% at z=0.1, 0.3 and 0.6, respectively.

\section{BCG radio properties and interaction with the ICM
  as a function of cosmic epoch}
\label{feedback.sec}
\vspace{-0.1in}

The radio-mode feedback is expected to be efficient in any cool core,
in order to prevent complete cooling and provide a smooth temperature
profile in the ICM as observed \citep{McNamara_2006, Rafferty_2008}.
The presence of a radio galaxy in the BCG of a cluster is predicted to
be generally associated to cavities in the ICM, as a result of the
interaction between the population of relativistic electrons and the
thermal electrons of the diffuse, X-ray emitting plasma.  The study of
cavities in the ICM inflated by relativistic electrons has been the
subject of several studies aimed at understanding the energetics of
the feedback processes, and the mechanism by which the mechanical
energy is transferred to the ICM.  The energetics can be estimated
thanks to the enthalpy ($4 P V$ for relativistic plasma) of the
bubbles \citep{Birzan_2004, Birzan_2008}, while the heating mechanism
is still strongly debated.  This mechanism is important also because
it is thought to be responsible for the heating of the gas at
galactic scales, efficiently quenching the star formation processes
and therefore setting the stellar mass scales and colors for massive
galaxies \citep{Croton_2006}.  In addition, strong radio sources can
be the beacon of massive protoclusters at redshift 2 and larger
\citep[see, e.g.,][]{Miley_2006}.  Therefore, radio selection of
clusters and protoclusters may be an efficient method to find the
progenitor of massive clusters at $z>2$, where the red sequence of the
cluster galaxies may not be fully established, and the ICM is not
shining in the X-ray band yet, indicating an incomplete virialization
\citep{Chiaberge_2010}.

\begin{figure*}[t]
\hspace{1.5cm}
\parbox{0.75\textwidth}{
\centerline{\includegraphics[scale=0.5]{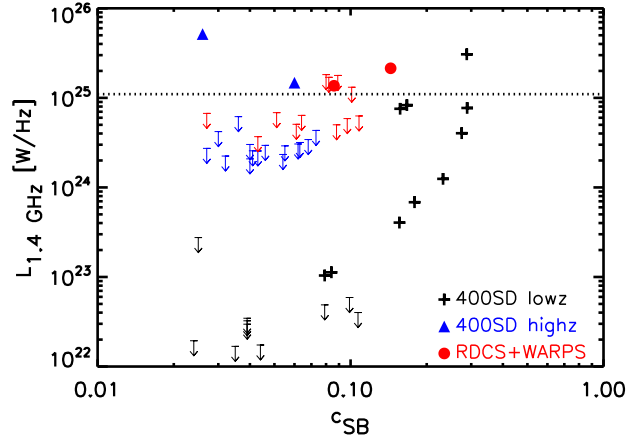}}
}
\vspace{-0.3cm}
\caption{\label{santos.fig} Relation between the radio luminosity of
  sources in close proximity to the cluster cores, $L_{1.4}$ GHz, and
  the surface brightness concentration, $C_{SB}$. Arrows refer to
  upper limits of non-detections. The dotted line at $1.1 \times
  10^{25}$ W Hz$^{-1}$ marks the luminosity limit at $z = 1$ for NVSS
  sources, corresponding to the flux limit of 2.5 mJy
  \citep[from][]{Santos_2010}.  }
\end{figure*}

Tracing the feedback in galaxy clusters in the radio band consists of
detecting the presence of radio AGN in the central galaxy, measuring
its total power and resolving the radio emission in the radio lobes
filling the cavities in the ICM.  These studies have been very
successful for local targets, where deep radio and X-ray data are both
available, but they become increasingly difficult at higher redshifts.
Recently, a few studies focused on high-redshift clusters and on the
evolution of their properties with cosmic time.  In particular, it has
been realized that the evolution of the cool-core phenomenon is
mirrored by the evolution of the feedback process itself. A simple
correlation of the radio power and the cool-core strength (quantified
through the concentration parameter $C_{SB}$) has been shown for low
and medium redshift clusters \citep[see Figure \ref{santos.fig} right
panel,][]{Santos_2010}.  It is observed that the luminosity of radio
galaxies in the center of cool-core cluster have radio powers ranging
from $\sim 10^{23}$ W Hz$^{-1}$ to $10^{25}$ W Hz$^{-1}$.  Ongoing
JVLA programs (PI P. Tozzi) are being carried out to target the
majority of the high redshift clusters with X-ray data known to date,
to extend this relation up to $z\sim 1$, and to perform a systematic
investigations of the radio-mode feedback in a representative sample
of massive clusters in the CLASH sample \citep{Postman_2012}.  A
comprehensive study of the radio properties of BCGs in X-ray selected
galaxy clusters is in preparation (Hogan et al., submitted), in which
multi-frequency radio observations allow one to investigate how the
BCG properties vary with cluster environmental factors.

\begin{figure}[t]
\centering 
\includegraphics[width=.48\textwidth,origin=c,angle=0]{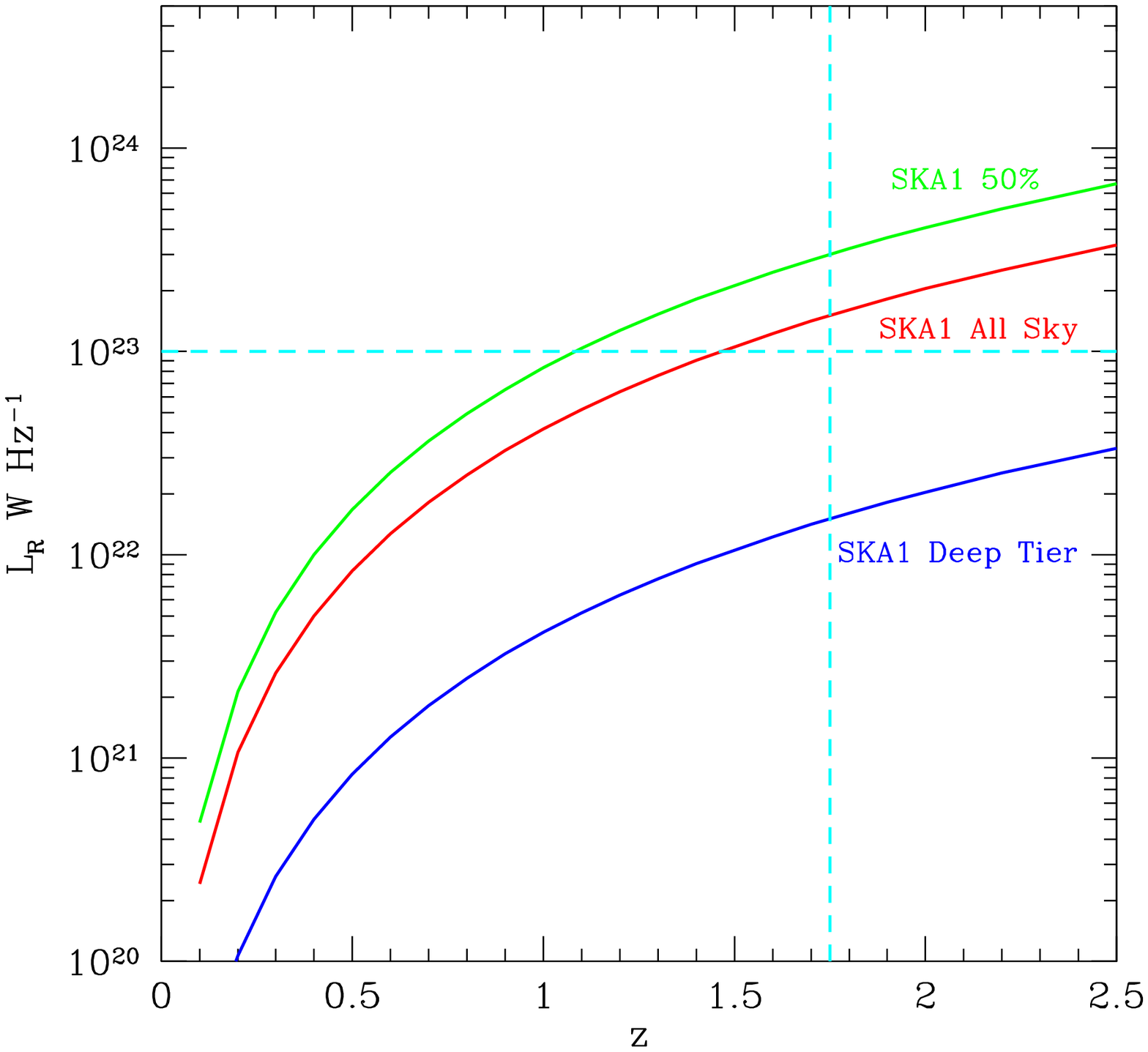}
\hspace{0.4cm}
\includegraphics[width=.48\textwidth,origin=c,angle=0]{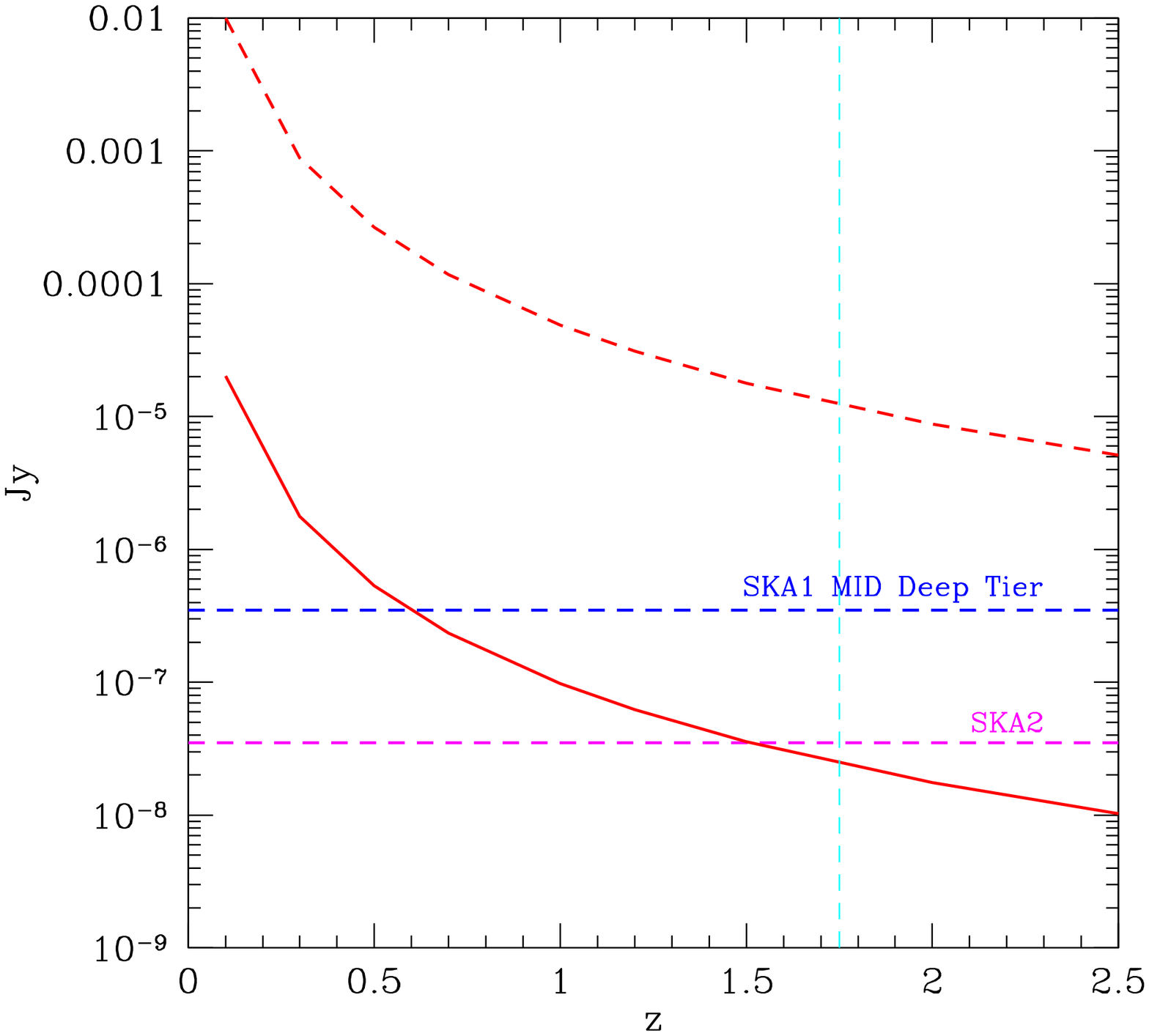}
\vspace{-1.cm}
\caption{\label{fluxlim.fig} {\it Left panel:} Radio power at 1.4 GHz
  detectable with a sensitivity of 1-10-20 $\mu Jy$ (defined as $5
  \times$ rms noise) are shown in blue, red, and green lines,
  respectively, as a function of redshift.  Green corresponds to early
  SKA, red to SKA1 All Sky, and blue to SKA1 Deep Tier.  The
  horizontal dashed line shows the minimum radio luminosity currently
  measured in a radio galaxy hosting a cool core in local clusters.
  {\it Right panel:} the dashed red line shows the flux density per
  beam with a resolution of 0.75 arcsec for a bubble with radio power
  of $10^{24}$ W Hz$^{-1}$ as observed in the cluster RBS~797.  The
  dashed horizontal blue line marks the sensitivity of Deep Tier (2000
  hrs) reference surveys with SKA1-MID at the same angular resolution.
  The solid red line shows the flux density per beam with a resolution
  of 0.3 arcsec for a bubble with radio power of $2 \times 10^{21}$ W
  Hz$^{-1}$ as observed in the group HCG~62.  The magenta dashed line
  shows the sensitivity reacheble with SKA2.  In both panels the
  vertical dashed lines mark the highest redshift where virialized,
  X-ray emitting clusters are currently detected.  \vspace{-0.1in}}
\end{figure}

From a simple visual inspection of Figure \ref{santos.fig}, we
conclude that we need to detect every radio galaxy at least down to
$10^{23}$ W Hz$^{-1}$ (at 1.4 GHz) in order to explore the same
dynamical range at high redshift.  In the left panel of Figure
\ref{fluxlim.fig} we show the radio power in unresolved sources
corresponding to detection limits of 20, 10 and 1 $\mu$Jy from top to
bottom.  Assuming that typically a detection requires a signal to
noise ratio $S/N \sim 5$, these values correspond to 4 $\mu$Jy,
2$\mu$Jy and $0.2 \mu$Jy per beam rms, which will be achieved by SKA1
(50\% sensitivity), SKA1 All Sky and SKA1 Deep Tier, respectively.
With early SKA1 (50\% sensitivity) it will be possible to detect all
the feedback process involving radio galaxies with power larger than
$10^{23}$ W Hz$^{-1}$ up to $z \leq 1$.  We find that sources above
$10^{23}$ W Hz$^{-1}$ can be detected up to redshift $z\sim 1.7$ (the
largest redshift where virialized galaxy clusters are currently found)
and beyond with a sensitivity of 10 $\mu$Jy per beam, which is easily
achieved by SKA1-MID on the entire sky (3$\pi$) with a subarcsec
resolution.  At this sensitivity, confusion is avoided for a beam FWHM
smaller than 4 arcsec \citep[see][]{Prandoni-SKA}.  A ten time lower
detection threshold, corresponding to an rms of $0.2 \mu$Jy per beam
(blue line in Figure \ref{fluxlim.fig}, left panel) is reached in the
Deep Tier (2000 hrs) reference surveys (over about 10-30 deg$^2$) with
SKA1-MID.  This sensitivity is obtained at an angular resolution below
1 arcsec, thus easily allowing a complete census of the radio
properties of the BCG at any redshift down to a total power of
$10^{22}$ W Hz$^{-1}$ at 1.4 GHz.  In this case, even in the presence
of a strong negative evolution with redshift, all the BCG in massive
clusters with cool core and feedback activity will be detected and
characterized.
Considering that on the entire sky there are potentially $4\times
10^4$ massive clusters at $z>0.5$ which are within the reach of Athena
or any X-ray survey mission with an average resolution of $\leq 10$
arcsec (like the Wide Field X-ray Telescope\footnote{see
  www.wfxt.eu}), the number of galaxies with radio activity in cool
cores at $z>0.5$ can be as high as $\sim 2 \times 10^4$, assuming
little or no evolution in the population of cool core clusters, or few
thousands assuming a strong evolution by a factor of 10.
At present, there is large uncertainty on the evolution of cool cores
\citep{Santos_2010,McDonald_2013}, and only a strong synergy of SKA1
with deep and wide X-ray data will answer to this question.  Clearly
SKA, when completed, will allow one to investigate the presence of
radio galaxies also in groups at high redshift, a completely new
window for which it is impossible to make predictions given the
present knowledge of the feedback processes.  SKA therefore offers a
unique opportunity to constrain the duty cycle of radio galaxies in
groups and cluster of galaxies across the cosmic epochs.  Another
relevant aspect to consider is that radio BCGs often show variabiliy
and a high complexity in their radio spectra (Hogan et al.,
submitted), with $\sim$80\% of them showing a
flat-spectrum/self-absorbed core.  These flatter/active components
will often be distinguishable only at frequencies larger than 5 GHz,
making it desirable to have coverage at these frequencies.

A further major breakthrough in the investigation of radio-mode AGN
feedback at high redshift is the detection and characterization of the
radio lobes produced by the relativistic electrons which are
responsible for carving the cavities in the ICM clearly observed
in X-ray local clusters.  If we use as a reference the bubble size in
the two objects shown in Figure \ref{bubbles.fig}, we can 
  directly compute the typical angular size of the bubbles as a
function of the redshift.  For the group HCG~62 ($z=0.0137$) the
average size for a single bubble, assuming a spherical shape, is 5
kpc, while for the medium-mass cluster RBS~797 ($z=0.35$) is 12 kpc.
If we focus on targets at $z>1$ the angular size turns out to be about
0.6 and 1.5 arcsec, respectively, with a weak dependence on the
redshift.  For these objects we measure a total radio power at 1.4 GHz
of $2.0 \times 10^{21}$ W Hz$^{-1}$ and $10^{24}$ W Hz$^{-1}$ for a
single cavity in HCG~62 and RBS~797, respectively.  We conservatively
assume that a beam FWHM of half the size of the bubble is sufficient
to resolve it.  Therefore, we compute the radio flux density at 1.4
GHz from a single cavity as a function of redshift, assuming a
spectral index $\alpha \sim 1$, and a beam FWHM of 0.3 arcsec and 0.75
arcsec for the reference cases of HCG~62 and RBS~797, respectively.
We show in Figure \ref{fluxlim.fig} (right panel) that the signal from
a single cavity in medium and large clusters is always well above the
sensitivity of Deep Tier (2000 hrs) reference surveys with SKA1-MID,
which is computed as $5\times 70$ nJy per beam rms
\citep[see][]{Prandoni-SKA}.  At this resolution, we are also well
above the confusion level.  Therefore the detection and
characterization of typical cavities in medium or large mass cluster
is within the reach of SKA1-MID surveys at any relevant redshift.

On the other hand, to detect a single cavity in a small group a beam
FWHM of 0.3 arcsec is required, and given the much lower radio power
of $\sim 2.0 \times 10^{21}$ W Hz$^{-1}$, the expected radio flux
density in the beam rapidly falls below the sensitivity limit of
SKA1-MID (see red lines in Figure \ref{fluxlim.fig}, right panel).
According to the current specifics, SKA1-MID allows detection and
characterization of radio bubbles only up to $z\sim 0.5$.  If a $\sim
10$ times better sensitivity can be reached with SKA2 at the
resolution of 0.3 arcsec, the investigation of the radio-mode feedback
becomes feasible also in groups up to $z\sim 1.3$.  We also note that
the fraction of BCGs that show clear radio lobes in local X-ray
selected clusters is about 10\% (Hogan et al., submitted).  This
leaves room for several thousands of clusters in the entire sky
potentially hosting bubbles and cavities, but it may also indicate
that most of the activity we actually observe around radio galaxies is
either the aftermath of lobe creation and dissipation, or, more
likely, low power outbursts that are confined within a few hundreds pc
of the core. In this case at high redshift the full resolution SKA
images will likely appear unresolved or amorphous. Nevertheless, the
subarcsec angular resolution provided by the long baselines are
essential to investigate the feedback process in details in relatively
low and medium redshift clusters.

To summarize, our preliminary feasibility study shows that the
radio-mode feedback will be unveiled practically at any level in
clusters up to $z \sim 1.7$, the maximum redshift where virialized
clusters have been detected in the X-ray band so far.  The same study
will be possible also in high-redshift groups if SKA2 will improve the
sensitivity by an order of magnitude at the same angular resolution.
When SKA will be fully operational, the knowledge of distant galaxy
clusters will be largely but unpredictably changed.  A large number of
optically and IR selected clusters will be available thanks to the
forthcoming surveys of the Euclid\footnote{http://sci.esa.int/euclid/}
satellite and of the LSST project\footnote{http://www.lsst.org/lsst/}.
Thanks to the ICM X-ray characterization of the cluster cores
achievable with the Athena mission, these can be directly combined
with the radio data to search for cavities.  However, an efficient
synergy with SKA will depend on the on-axis angular resolution of the
X-ray telescope, which should be of the order of $1$ arcsec.  In the
field of high-z clusters, as in many others, the advent of SKA will
provide access to incredibly deep and high quality data, which are
hardly matched by any planned of foreseen facility in other wavebands.

\bibliographystyle{apj_short_etal.bst}
\bibliography{bibliography-gitti.bib}

\end{document}